\documentclass[%
 aip,
 amsmath,amssymb,
 reprint,%
]{revtex4-1}

\usepackage{graphicx}
\usepackage{dcolumn}
\usepackage{bm}

\usepackage[utf8]{inputenc}
\usepackage[T1]{fontenc}
\usepackage{mathptmx}
\usepackage{etoolbox}
\usepackage{lipsum}

\usepackage{siunitx}
\usepackage{mhchem}
\usepackage{makecell}
\usepackage{booktabs}

\DeclareSIUnit{\atpercent}{at.\%}
\DeclareSIUnit{\rpm}{rpm}
\DeclareSIUnit\torr{Torr}
\DeclareSIUnit\sccm{sccm}

\makeatletter
\def\@email#1#2{%
 \endgroup
 \patchcmd{\titleblock@produce}
  {\frontmatter@RRAPformat}
  {\frontmatter@RRAPformat{\produce@RRAP{*#1\href{mailto:#2}{#2}}}\frontmatter@RRAPformat}
  {}{}
}%
\makeatother
\begin{document}

\preprint{AIP/123-QED}

\title[]{Composition Effects on Ni/Al Reactive Multilayers: A Comprehensive Study of Mechanical Properties, Reaction Dynamics and Phase Evolution}
\author{N. Toncich}
 \affiliation{Laboratory for Nanometallurgy, Department of Materials, ETH Zurich}
 \email{nensi.toncich@mat.ethz.ch}
 \author{F. Schwarz}
 \affiliation{Laboratory for Nanometallurgy, Department of Materials, ETH Zurich}
\author{R.A. Gallivan}
 \affiliation{Laboratory for Nanometallurgy, Department of Materials, ETH Zurich}
 \author{J. Gillon}
 \affiliation{Laboratory for Nanometallurgy, Department of Materials, ETH Zurich}
\author{R. Spolenak}
 \affiliation{Laboratory for Nanometallurgy, Department of Materials, ETH Zurich}

\date{\today}

\begin{abstract}
Ni/Al reactive multilayers are promising materials for applications requiring controlled local energy release and superior mechanical performance. This study systematically investigates the impact of compositional variations, ranging from 30 to \qty{70}{\atpercent} Ni, and bilayer thicknesses (\qty{30}{\nano\meter} and \qty{50}{\nano\meter}) on the mechanical properties and reaction dynamics of Ni/Al multilayers. Multilayers with varying Ni-to-Al ratios were fabricated and subjected to instrumented nanoindentation testing to evaluate hardness and elastic modulus. Combustion experiments, conducted on dogbone-shaped multilayers deposited onto silicon wafers with thermal barrier coatings, characterized the reaction front's speed, temperature, and the resulting phases. The findings revealed that composition variations within this range enable precise tuning of reaction speed and temperature without significant changes in mechanical properties, while deviations in modulus and hardness at higher nickel concentrations suggest microstructural influences. Notably, phase formation in Al-rich samples deviated from equilibrium predictions, highlighting the role of kinetic factors, such as diffusion and rapid quenching, in driving non-adiabatic processes during phase evolution. Molecular dynamics simulations provided complementary atomistic insights into mechanical responses and reaction kinetics, bridging experimental observations with theoretical predictions. This integrated approach advances the understanding of Ni/Al multilayers, offering a framework for optimizing their composition and structural design to achieve tailored performance for application-specific requirements.
\end{abstract}

\maketitle

\section{\label{sec:level1}INTRODUCTION}

Reactive multilayers (RMs) are energetic materials consisting of alternating thin films from at least two different components, characterized by both a large negative enthalpy of mixing and a high adiabatic reaction temperature~\cite{adams2015reactive}. External stimuli such as heating, electrical sparks, mechanical impact, or laser irradiation can trigger layer intermixing which leads to a highly exothermic reaction. If the heat generated exceeds the heat dissipated in the surrounding environment, the reaction, driven by the low free energy of the products, can propagate in a self-sustained manner~\cite{fritz2013thresholds,picard2008nanosecond,weihs2014fabrication}.

These systems demonstrate significant potential for innovation and are actively utilized in various applications due to their rapid, localized, and intense release of thermal energy. Their adaptability lies in the ability to precisely tailor key properties, such as bilayer thickness~\cite{knepper2009effect}, composition~\cite{danzi2019architecture}, and interface characteristics~\cite{schwarz2022influence}, to meet specific requirements. This customization allows for the optimization of mechanical properties, energy output, reaction speed, and thermal behavior, making these systems highly effective for applications including joining and bonding~\cite{wang2005effects,bourim2021investigation}, low-temperature synthesis~\cite{lee2014low}, igniters~\cite{zeng2019mechanism}, power sources~\cite{guidotti2006thermally}, and self-healing of thin films~\cite{danzi2019rapid}. Thus, ongoing research is dedicated to understanding control mechanisms, aiming to fine-tune the properties and behaviors involved~\cite{schwarz2024qualitatively}.

Among these systems, Nickel/Aluminum (Ni/Al) RMs have attracted significant attention, emerging as the most extensively studied multilayer systems both experimentally and through Molecular Dynamics (MD) simulations. The studies demonstrate that reaction temperature and speed can be manipulated by varying the bilayer thickness~\cite{knepper2009effect}, the architecture of the system and substrate~\cite{fritz2011enabling,sauni2024controlling}, as well as the composition and the microstructure of the layers~\cite{dyer1995synthesis,danzi2019architecture,schwarz2022md}. 
Additionally, adjusting the stoichiometry impacts various system properties, including reaction front behavior~\cite{gavens2000effect,sen2017synthesis} and mechanical strength~\cite{schwarz2024investigating}. Stoichiometry also affects the heat of reaction, which consists of two main components: the heat of mixing, which is the energy released or absorbed when the reactants combine, and the heat of crystallization, which arises from the formation of a new phase as the reaction proceeds~\cite{weihs2014fabrication,schwarz2022influence}. Although Ni/Al RMs are among the most studied systems, research has predominantly focused on a limited range of compositions, such as Ni/3Al~\cite{ma1990self}, 2Ni/3Al~\cite{gavens2000effect}, Ni/Al~\cite{dyer1995synthesis, rogachev2012self, ma2018microstructures}, 3Ni/Al~\cite{dyer1995synthesis}, and 3Ni/2Al~\cite{sauni2023microstructural}.  Recent studies have expanded the investigation of Ni/Al systems to include compositions with nickel content ranging from 5 to \qty{25}{\atpercent} Ni~\cite{neuhauser2022effect}.

Research on Ni/Al RMs reveal composition and experimental parameters to have substantial influence on reaction kinetics and phase formation. For the 3Ni/Al system, Dyer and Munir~\cite{dyer1995synthesis} reported reaction speeds ranging from 0.08 to \qty{0.25}{\meter\per\second}, with the formation of \ce{Ni3Al} and traces of NiAl. Conversely, Ma \textit{et al.}~\cite{ma1990self} observed a significantly higher reaction speed of \qty{4}{\meter\per\second}, resulting exclusively in \ce{Ni3Al}. This disparity in reaction speeds can be attributed to several factors, including the presence of impurities in the source materials used for e-beam evaporation in Dyer and Munir's study and the use of thicker bilayer thicknesses, both of which could reduce reaction rates. Additionally, their study of the Ni/3Al system, which exhibited the same reaction speed, revealed the formation of \ce{NiAl3}, Al, and \ce{Ni2Al3}. In the case of the 2Ni/3Al system, reaction speeds were reported between 1.1 and \qty{10.1}{\meter\per\second}, with \ce{Ni2Al3} being the final phase~\cite{gavens2000effect, blobaum2003ni}. Van Herdeen \textit{et al.}~\cite{van1997metastable} also identified \ce{Ni2Al3}, with some traces of \ce{NiAl3}. The Ni/Al system with a 1:1 atomic ratio is among the most extensively studied, exhibiting reaction speeds from 0.6 to \qty{10}{\meter\per\second} and consistently forming the B2 NiAl phase~\cite{dyer1995synthesis, rogachev2012self, ma2018microstructures}. This wide range in reaction speeds can be attributed to differences in fabrication processes and parameters, which influence the microstructure and the extent of intermixing between Ni and Al. Variations in the intermixed region, often formed during deposition, can affect heat generation and dissipation~\cite{schwarz2022influence}. Additionally, differences in total system thickness may further contribute to the variability. The maximum observed combustion temperatures for this composition range from \qty{1500}{\degreeCelsius} to \qty{1700}{\degreeCelsius}~\cite{rogachev2014structure, danzi2019architecture, danzi2019thermal}. In contrast, the 3Ni/2Al ratio resulted in a lower combustion temperature of approximately \qty{1100}{\degreeCelsius}~\cite{sauni2023microstructural}. Furthermore, recent studies on samples with nickel concentrations below 25\% have demonstrated that the formation of phases such as \ce{NiAl3} and \ce{Ni2Al9} is influenced both by the nickel content and the heating rate~\cite{neuhauser2022effect}.

\begin{figure*}
\includegraphics{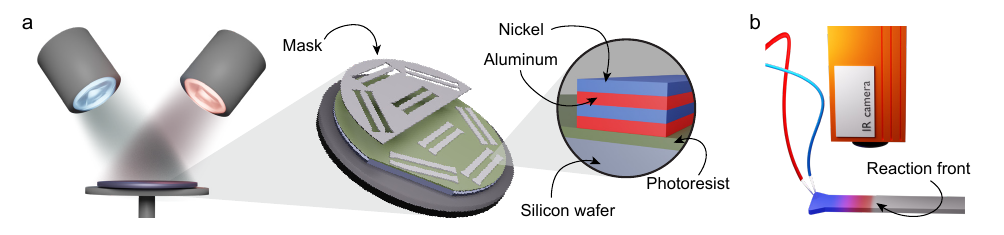}
\caption{\label{fig:schem}Schematic representation of the experimental setup. (a) Ni/Al multilayers were deposited by magnetron sputtering onto a Si wafer coated with a thermally insulating photoresist. To prevent oxidation, aluminum was sputtered as the first layer, followed by nickel as the final layer. Dogbone-shaped multilayer systems were obtained using a steel stencil mask. (b) The dogbone samples were ignited with an electrical spark, and the reaction was recorded by a high-speed infrared camera positioned above the multilayers.}
\end{figure*}
 
While research has primarily focused on reaction of these reactive multilayers, mechanical properties are a critical consideration for applications where multilayers are integrated into systems, but not immediately triggered. The mechanical properties of metallic multilayers, particularly non-elastic properties such as strength and hardness, often deviate from the rule of mixtures, necessitating a more detailed investigation of their behavior~\cite{ebrahimi2024high}.
Specifically, the mechanical properties of nanolaminates, such as hardness and strength, are strongly influenced by the thickness and microstructure of the layers. Traditionally, three regimes describe this behavior: the Hall–Petch regime at layer heights from several hundred to below \qty{100}{\nano\meter}, where strength increases with decreasing bilayer height \(B\) due to dislocation pileup; the confined layer slip (CLS) regime, where further strength increase occurs due to dislocation confinement, following a \(\ln(B)/B\) scaling; and the inverse Hall–Petch regime at very small heights, where strength decreases, attributed to interface effects, dislocation crossing, and reduced dislocation stability. The specific thickness at which this regime occurs depends on the deposition method, as different techniques influence factors such as interdiffusion and grain size~\cite{hall1951deformation,petch1953cleavage,misra2001deformation,misra2005length,jian2022role,ren2011theoretical}. However, recent studies challenge the existence of a distinct Hall–Petch regime, suggesting instead that the CLS regime governs strengthening across all scales, with a universal \(\ln(B)/B\) scaling~\cite{matthews1970accommodation,RIP_Hall_Petch,schwarz2024investigating}.

Nasim \textit{et al.} studied the mechanical properties of Ni/Al multilayers through micropillar compression and nanoindentation, focusing on the effect of bilayer thickness, where the individual layers were kept at equal thickness~\cite{Exp_comp_AlNi_multilayer,Exp_indent_AlNi_multilayer}. The study found that Ni/Al multilayers show increased hardness and strength as the bilayer thickness decreases, reaching maximum values at the individual layer thickness of \qty{10}{nm}. The elastic modulus is highest at \qty{20}{nm}, while below these thicknesses, mechanical properties deteriorate due to plastic deformation and broken interfaces. Through stoichiometry variation, Shen \textit{et al.}~\cite{shen2023mechanical} observed a significant hardness and yield strength increase with decreasing Al layer thickness, as thickness drops to \( \leq \) \qty{200}{nm}. The deformation is primarily influenced by the constrained plasticity of Al layers, and the confined layer slip model effectively describes the observed strengthening, driven by confined dislocation glide in the Al layers. These studies demonstrate that the mechanical properties of Ni/Al multilayers are strongly influenced by both bilayer thickness and stoichiometry, with experimental compositions spanning a wide range of Ni atomic percentages~\cite{shen2023mechanical,Exp_indent_AlNi_multilayer,Exp_comp_AlNi_multilayer,roll_bond_AlNi_exp,AlNi_exp_ind_wear}; however, the bilayer height was not held constant during these variations.

Furthermore, MD simulation studies explored the mechanical properties of equiatomic and equivolumetric Ni/Al systems, finding that nanocrystalline Ni layers closely matched experimental results. This highlights the importance of accurate microstructure characterization. Hardening followed \( B^{-1/2} \) for bilayer heights $\geq$ \qty{50}{nm} and \(\ln(B)/B\) scaling for $\geq$ \qty{15}{nm} which is consistent with the confined layer slip model~\cite{schwarz2024_comp}. In another MD simulation study, it was found that the premixed interlayer only significantly impacted mechanical properties at the smallest bilayer height of \qty{5}{nm}, while the top layer material (Ni or Al) had a markedly stronger influence on hardness and stiffness~\cite{schwarz_nanoind}.

These findings underscore the complex interplay between composition, reaction conditions, and phase development in Ni/Al RMs. Variations in key parameters such as fabrication processes, bilayer thickness, number of bilayers, and overall thickness, significantly influence mechanical properties, reaction kinetics, and phase outcomes. Notably, discrepancies in the reported results, particularly in reaction speed, temperature, and resulting phase, highlight inconsistencies across different studies. To address these gaps, we propose a systematic investigation using a consistent methodology and experimental setup. This approach aims to resolve existing inconsistencies and establish a comprehensive understanding of the underlying mechanisms governing Ni/Al RMs.

Therefore, in this work we explore the impact of composition variation in Ni/Al RMs on their mechanical and reaction properties, supported by MD simulations for atomistic insights. To study the effect of composition variation at a constant bilayer height, nine compositions ranging from 30 to \qty{70}{\atpercent} nickel were experimentally investigated at two bilayer thicknesses (\qty{30}{\nano\meter} and \qty{50}{\nano\meter}). In this work, we demonstrate that composition variation is a tool that allows to vary both mechanical and reaction properties of RMs over a wide range. Careful investigation allows us to quantify the microstructure and reaction products, shedding light on the underlying mechanisms.

\section{\label{sec:level2}Materials \& Methods}

\subsection{Reactive Multilayer Systems Deposition}
\label{subsec:sputtering}

Ni/Al multilayers were deposited using a commercial magnetron sputtering system (PVD Product, Inc.) on a silicon wafer coated with a thermally insulating photoresist. The silicon wafer was cleaned with acetone, isopropanol and subsequently rinsed with water. The positive photoresist AZ 4533 was spin coated on the Si wafer at \qty{4000}{\rpm} for \qty{50}{\second}~\cite{danzi2019thermal}. Ni (MaTeck GmbH, 99.99\%) and Al (MaTeck GmbH, 99.99\%) layers were deposited by a DC current at 100 and \qty{250}{\watt}, respectively, under a \qty{3}{\milli\torr} Ar pressure at \qty{50}{\sccm}. The base pressure was kept at \qty{e-7}{\torr} for all depositions. The substrate rotation was set at \qty{30}{\rpm} to ensure the uniformity of the layer thickness and its temperature was kept below \SI{40}{\degreeCelsius}. The composition and bilayer thickness of various Ni/Al multilayer systems were adjusted by varying the respective thicknesses of the Ni and Al layers. This was accomplished through precise control of the sputtering time. In all depositions, Al was the first layer to be sputtered, and Ni was the last. A total of 50 bilayers were deposited. A steel mask with dog bone-shaped holes was attached to the substrate to impart this shape to the multilayers, allowing for multiple ignitions (FIG.~\ref{fig:schem}(a)). To ensure the deposition of flat layers throughout the entire sample, the substrate temperature was maintained at ambient conditions. This was achieved by sputtering 5 bilayers sequentially, followed by an intermission before resuming the deposition process. This controlled approach was key to achieve uniformity in the multilayer structure. For the various measurements, the multilayers were deposited on different substrates: Si wafer chips for XRD analysis of the as-deposited samples, glass slides for deposition calibration (employing an AFM alpha300 RA), and Si wafers coated with photoresist for the remaining characterizations. A list of the sputtered compositions is shown in Table \ref{tab:table1}.
The samples were named based on their composition; for example, the sample 30Ni70Al indicates a composition of \qty{30}{\atpercent} Ni and \qty{70}{\atpercent} Al.

\begin{table}
\caption{\label{tab:table1}Experimentally studied Ni/Al compositions and their nominal layer thicknesses, along with the as-deposited mechanical properties.}
\begin{ruledtabular}
\begin{tabular}{cccccccc}
\makecell{$Ni$ \\ (at.\%)} & \makecell{$B$ \\ (\si{\nano\meter})} & \makecell{$h_{\mathrm{Ni}}$ \\ (\si{\nano\meter})} & \makecell{$h_{\mathrm{Al}}$ \\ (\si{\nano\meter})} & \makecell{$E$ \\ (\si{\giga\pascal})} & \makecell{$H$ \\ (\si{\giga\pascal})} \\
\hline
30 & \makecell{$30$ \\ $50$} & \makecell{$6.6$ \\ $11.0$} & \makecell{$23.4$ \\ $39.0$} &\makecell{$102.9  \pm 0.3$ \\ $104.8 \pm 1.8$ } & \makecell{$4.08 \pm 0.01$ \\ $4.32 \pm 0.03$} \\
\hline
35 & \makecell{$30$ \\ $50$} & \makecell{$7.9$ \\ $13.1$} & \makecell{$22.1$ \\ $36.9$} & \makecell{$107.5  \pm 1.6$ \\ $113.2 \pm 1.2$} & \makecell{$4.31 \pm 0.03$ \\ $4.47 \pm 0.00$} \\
\hline
40 & \makecell{$30$ \\ $50$} & \makecell{$9.2$ \\ $15.3$} & \makecell{$20.8$ \\ $34.7$} & \makecell{$113.8 \pm 0.6$ \\ $121.4 \pm 1.4$} & \makecell{$4.84 \pm 0.01$ \\ $4.80 \pm 0.02$} \\
\hline
45 & \makecell{$30$ \\ $50$} & \makecell{$10.5$ \\ $17.5$} & \makecell{$19.5$ \\ $32.5$} & \makecell{$125.2 \pm 1.6$ \\ $126.2 \pm 0.2$} & \makecell{$4.94 \pm 0.01$ \\ $5.09 \pm 0.01$} \\
\hline
50 & \makecell{$30$ \\ $50$} & \makecell{$11.9$ \\ $19.9$} & \makecell{$18.1$ \\ $30.1$} & \makecell{$134.4 \pm 0.7$ \\ $116.9 \pm 3.1$} & \makecell{$5.22 \pm 0.02$ \\ $4.68 \pm 0.01$} \\
\hline
55 & \makecell{$30$ \\ $50$} & \makecell{$13.4$ \\ $22.3$} & \makecell{$16.6$ \\ $27.7$} & \makecell{$136.9 \pm 4.2$ \\ $121.7 \pm 0.6$} & \makecell{$5.34\pm 0.03$ \\ $4.97 \pm 0.01$} \\
\hline
60 & \makecell{$30$ \\ $50$} & \makecell{$14.9$ \\ $24.9$} & \makecell{$15.1$ \\ $25.1$} & \makecell{$147.3 \pm 2.4$ \\ $131.1 \pm 1.9$} & \makecell{$5.62 \pm0.02$ \\ $5.32 \pm 0.01$} \\
\hline
65 & \makecell{$30$ \\ $50$} & \makecell{$16.5$ \\ $27.5$} & \makecell{$13.5$ \\ $22.5$} & \makecell{$148.4 \pm 1.4$ \\ $143.5 \pm 0.7$} & \makecell{$5.79 \pm 0.02$ \\ $5.76 \pm 0.00$} \\
\hline
70 & \makecell{$30$ \\ $50$} & \makecell{$18.2$ \\ $30.3$} & \makecell{$11.8$ \\ $19.7$} & \makecell{$145.9 \pm 1.8$ \\ $140.2 \pm 1.0$} & \makecell{$5.93 \pm 0.02$ \\ $5.77 \pm 0.03$} \\
\hline
\end{tabular}
\end{ruledtabular}
\end{table}

\subsection{Mechanical and Microstructural Characterization}

The mechanical properties of the as-deposited RM were measured using an iNano nanoindenter (Nanomechanics, Inc.) with an InForce50 actuator. All the samples were indented to a target load of \qty{6}{\milli\newton} using a diamond Berkovich tip (Synton-MDP). Indents were performed using the NanoBlitz$^{TM}$ program in a square grid of $10 x 10$ indents at 3 locations per condition with a spacing of \qty{3}{\micro\meter} between points.

XRD scans were performed in symmetrical Bragg-Brentano geometry, before and after ignition to determine the phase of the as-deposited systems and the phase transformation. A Panalytical X'Pert PRO MPD device was utilized, with a parallel beam geometry and copper $K_{\alpha}$ radiation.

To investigate the grain structure of both as-deposited and ignited samples, TEM lamellas were prepared from selected samples with a dual-beam Helios 5UX system (Thermo Fisher Scientific). Scanning transmission electron microscopy (STEM) and transmission electron microscopy (TEM) were conducted on a Talos F200X instrument (Thermo Fisher Scientific) operated at 200 kV, with additional selected area electron diffraction (SAED) performed.

\subsection{Ignition and High-Speed IR Imaging}
Ni/Al multilayer dogbone strips were ignited using a custom-built setup comprising an Agilent Technologies N6700B Low-Profile MPS mainframe as the power source, connected to micromanipulators equipped with tungsten probes. The reaction was initiated with a current pulse of \qty{2}{\volt} and \qty{10}{\milli\ampere}, following preliminary assessments that established these values as the threshold needed for ignition under ambient conditions~\cite{danzi2019thermal}.

The speed of the reaction front and the maximum temperature reached during the propagation of the self-sustained heat wave were tracked using a high-speed infrared camera (IRCAM Millenium 327k S/M) positioned above the multilayers (FIG.~\ref{fig:schem}(b)). Front propagation was recorded at a maximum frame rate of \qty{2.8}{\kilo\hertz}. The velocity was analyzed from the frames by measuring the position of the front over time. The temperature was calculated based on the emissivity of the uppermost material, i.e. Ni. For each sample, at least 3 ignitions were performed, and a minimum of 3 frames per ignition were considered for the measurements.

\subsection{MD Simulations}
The Molecular Dynamics (MD) simulations were conducted using the Large-scale Atomic/Molecular Massively Parallel Simulator (LAMMPS) package~\cite{LAMMPS}, employing the Embedded-Atom-Method (EAM) force field by Pun \textit{et al.}~\cite{Mishin_FF} to calculate the interatomic forces between individual atoms. This force-field has been shown to be suited for both qualitative simulations of the reaction front propagation~\cite{schwarz2021_microstructure,schwarz2022influence} as well as qualitative \& quantitative simulations of the mechanical properties~\cite{schwarz2024_comp,schwarz_nanoind} of Ni/Al reactive multilayers.

To study the reaction front propagation via MD simulations, systems of size $L\times$\qty{15}{\nano\meter}$\times$\qty{1.4}{\nano\meter}, with $L\geq$ \qty{400}{\nano\meter} were studied. The investigated compositions ranged from 33 $at\%$ Ni to 83 $at\%$ Ni. Both Al and Ni layers consisted of columnar grains of grain size equal to the individual layer heights, while at the interface, a premixed interlayer with a height of \qty{2}{\nano\meter} was inserted. The reaction speed was extracted as described in previous publications~\cite{schwarz2022influence,schwarz2022md,PhdThesis_Fabian_Schwarz}. The choice of a lower bilayer height than in experiments was due the absence of electron heat transport in MD simulations which leads to a lower reaction speed and thus at a bilayer height of \qty{50}{\nano\meter}, the multilayers do not ignite. The lower bound for the Ni content was chosen to include one system that does not ignite, while the upper bound was chosen such that the Al layer is not significantly smaller than the premixed interlayer. In the MD simulations, reaction front propagation beyond 70 $at\%$ Ni is observed, which is mainly attributed to the absence of heat loss in the MD simulations. For this reason, it is also difficult to extract a combustion temperature from the temperature distribution. While a quantitative comparison with experiment is not possible, a qualitative comparison can still lead to valuable insights on the atomistic level.

Uniaxial compression MD simulations of systems with [33, 40, 50, 60, 66] \qty{}{\atpercent} Ni were conducted to extract the mechanical properties, namely the Young's modulus, $E$, and the yield strength, $\sigma_y$. The systems were of dimensions $L_x\times L_y \times$ \qty{50}{\nano\meter}, with $L_x\geq$ \qty{30}{\nano\meter}, $L_y\geq$ \qty{30}{\nano\meter}. The Ni layers are made up of polycrystalline grains of size $d_{Ni}=$ min(\qty{10}{\nano\meter},$h_{Ni}$), where $h_{Ni}$ is the height of the Ni layer, while the Al layers are made up of columnar grains of size $d_{Al}=h_{Al}$, where $h_{Al}$ is the height of the Al layer. As perfect columnar grains (grain boundaries at 90 degrees to the interface) were shown to produce unrealistically high strength, an additional set of systems with a single crystal Al layer was studied. For the reaction front propagation simulations, a premixed interlayer with a height of \qty{2}{\nano\meter} was introduced at the Al-Ni interface. The Young's modulus and yield strength of such systems were shown to be in very good agreement with experimental results for the equivolume systems ($h_{Al}=h_{Ni}$) across a wide range of bilayer heights~\cite{schwarz2024_comp}. This publication also explains in detail the MD protocol and how the mechanical properties can be extracted from the MD trajectory.

\begin{figure}
\includegraphics{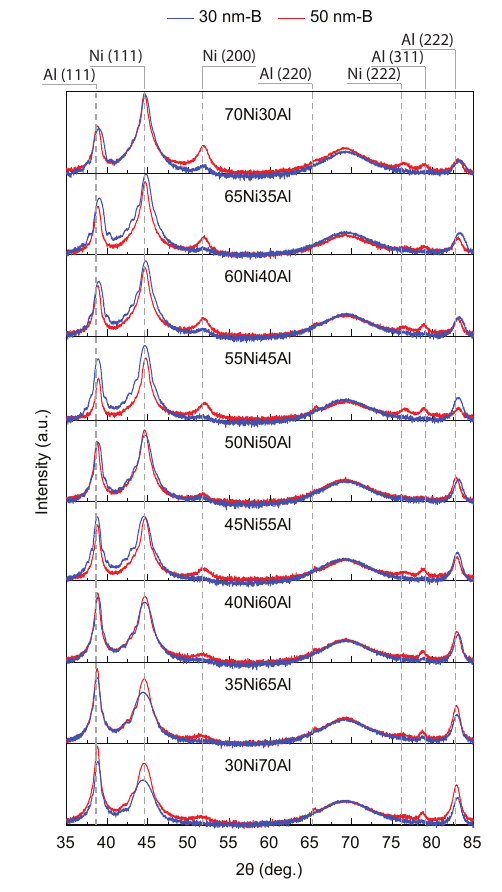}
\caption{\label{fig:XRDad}XRD spectra of the as-deposited Ni/Al RM with varying bilayer thicknesses and composition. The y-axis is logarithmically scaled.}
\end{figure}

\begin{figure*}
\includegraphics{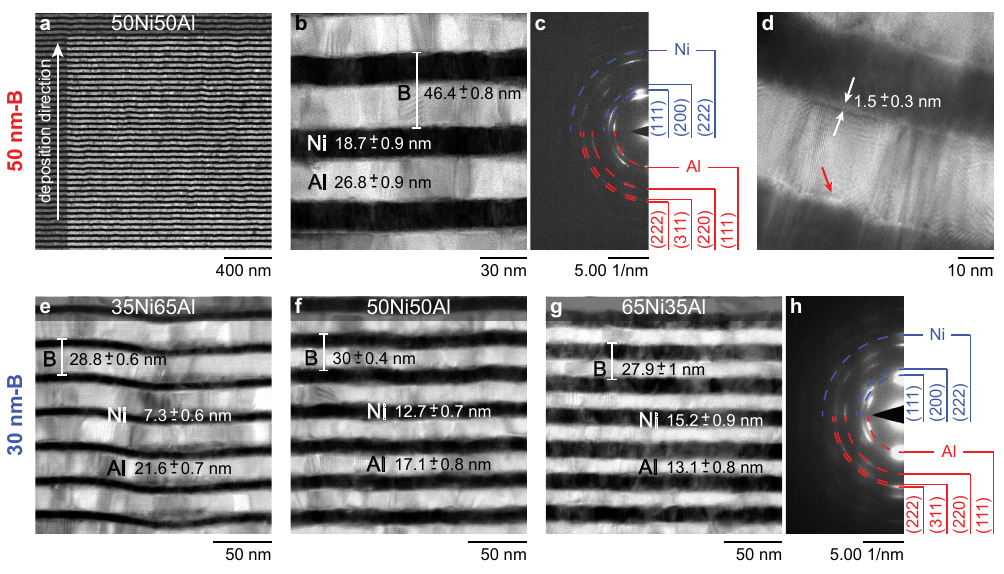}
\caption{\label{fig:tem_ad}BF-STEM and TEM images of selected as-deposited Ni/Al RM where the bright layers correspond to Al and the dark to Ni. (a) Low-magnified BF-STEM cross-section image of the 50Ni50Al RM with \qty{50}{\nano\meter}-B (b) High-magnified area from (a). (c) SAED patterns obtained from (b). (d) BF-TEM high-magnified image of the interface region. White arrows indicate the thickness of the interface region, while the red arrow point to the possible presence of a phase within the intermixed region. The (e-g) BF-STEM images of a \qty{30}{\nano\meter}-B thick 35Ni65Al, 50Ni50Al and 65Ni35Al, respectively. (h) SAED patterns obtained from (g).}
\end{figure*}

\section{Results}
\label{sec:results}

Two systems consisting of 30 and \qty{50}{\nano\meter}-thick bilayers (nm-B) were sputtered in a range of compositions from 30 to \qty{70}{\atpercent} Ni in steps of \qty{5}{\atpercent} leading to a total of 9 compositions per system. 
The fabricated Ni/Al dogbone-shaped systems are depicted in the schematics in FIG.~\ref{fig:schem}(a). This arrangement allowed for multiple tests of the same deposited sample.

\subsection{As-deposited sample analysis}

X-ray diffraction (XRD) patterns of the as-deposited systems are presented in FIG.~\ref{fig:XRDad}. All fabricated systems and compositions exhibited polycrystalline structures, with prominent peaks corresponding to the face-centered cubic (FCC) structures of Ni and Al. The predominant orientations observed were Al(111) and Ni(111), with notably high intensities. Variations in composition resulted in a decrease in intensity of Al-related peaks and an increase in Ni-related peaks as the Ni atomic percentage increased. A slight shift towards higher angles in the Al(111) peak was observed, possibly due to the interpenetration of Ni into the Al lattice~\cite{ramos2020effect}. The shift of the Ni peak is difficult to discern in FIG.~\ref{fig:XRDad}. Therefore, an analysis of the lattice constant is provided in the supplementary material (SFig.~1), where a shift toward higher (lower) values compared to the unstressed state is observed for the Ni (Al) layer as the composition approaches Al-rich (Ni-rich) regions. Alternatively, this effect may result from stress associated with the layer thickness or the formation of a metastable solid solution. The apparent large width of the Ni and Al peaks arises primarily from the use of a logarithmic scale, chosen to enhance the visibility of the main orientations. As the layer thickness decreases, peak broadening becomes more pronounced, attributed to a reduction in crystallite size (SFig.~2 in the supplementary material).
As the bilayer thickness decreases, continuous fluctuations, also known as satellites, become evident, reflecting interference effects typical of superlattices and multilayer structures. These satellite peaks evolve as a function of layer thickness and are more pronounced in systems with an ordered crystalline structure and sharp interfaces~\cite{lorenzin2022tensile}. No intermetallic phases were detected in the as-deposited samples, likely due to the substrate temperature being maintained below \SI{40}{\degreeCelsius} during sputtering. The broad peak observed at approximately 70° in the two-degree offset scan is attributed to thermally diffused scattering of the Si(400) reflection from the substrate.

\begin{figure*}
\includegraphics{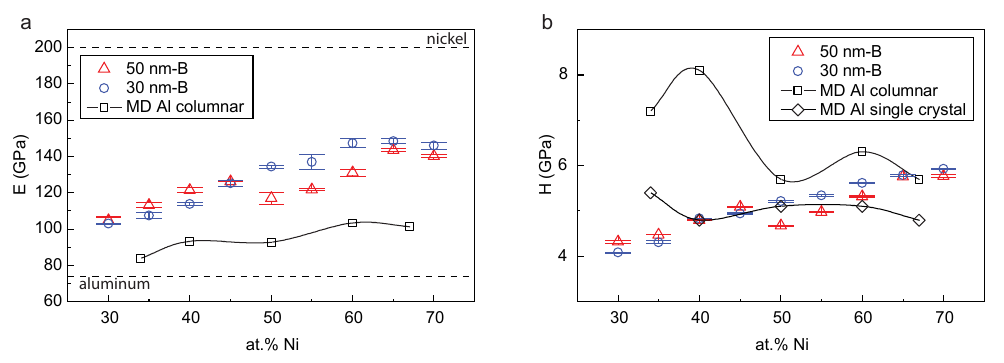}
\caption{\label{fig:indent}Experimental and MD simulation results for the mechanical properties of as-deposited 30 and \qty{50}{\nano\meter}-B systems. (a) Elastic modulus as a function of the \qty{}{\atpercent} Ni, with dashed lines indicating the modulus of pure Al~\cite{saha2002effects} and Ni~\cite{reddy2007mems}. (b) Hardness as a function of the \qty{}{\atpercent} Ni, including MD simulation results for both columnar Al grains and single crystal Al layers. Hardness values from the MD simulations were calculated by multiplying the extracted yield strength by a factor of 3.}
\end{figure*}

\begin{figure}
\includegraphics{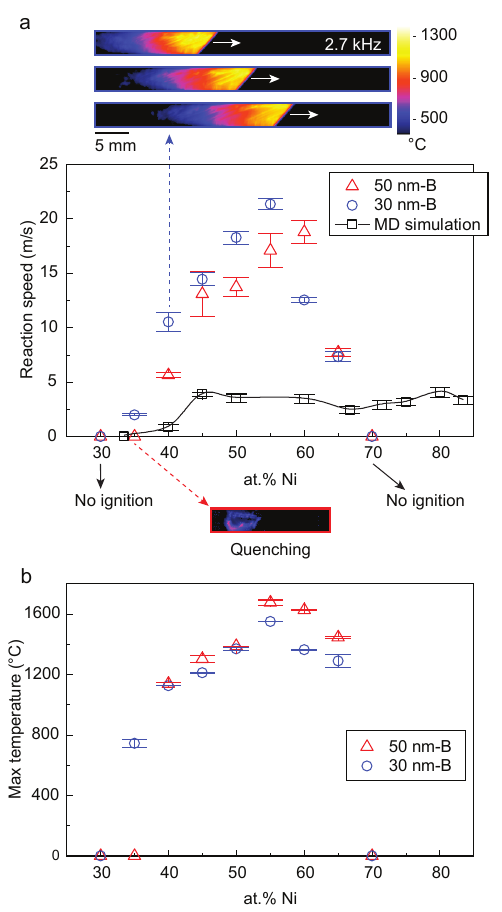}
\caption{\label{fig:speed} Reaction front analysis. (a-top) Frames depicting the self-sustained reaction of the 40Ni60Al with \qty{30}{\nano\meter}-B, acquired with the High-Speed IR Camera. (a-middle) Plot of the experimental and MD simulation results of the speed of the reaction as a function of the Ni concentration. (a-bottom) Single frame of the quenched reaction of the 35Ni65Al with \qty{50}{\nano\meter}-B. (b) Plot of the experimental result of the maximum reaction temperature as a function of the Ni concentration.}
\end{figure}

\begin{figure*}
\includegraphics{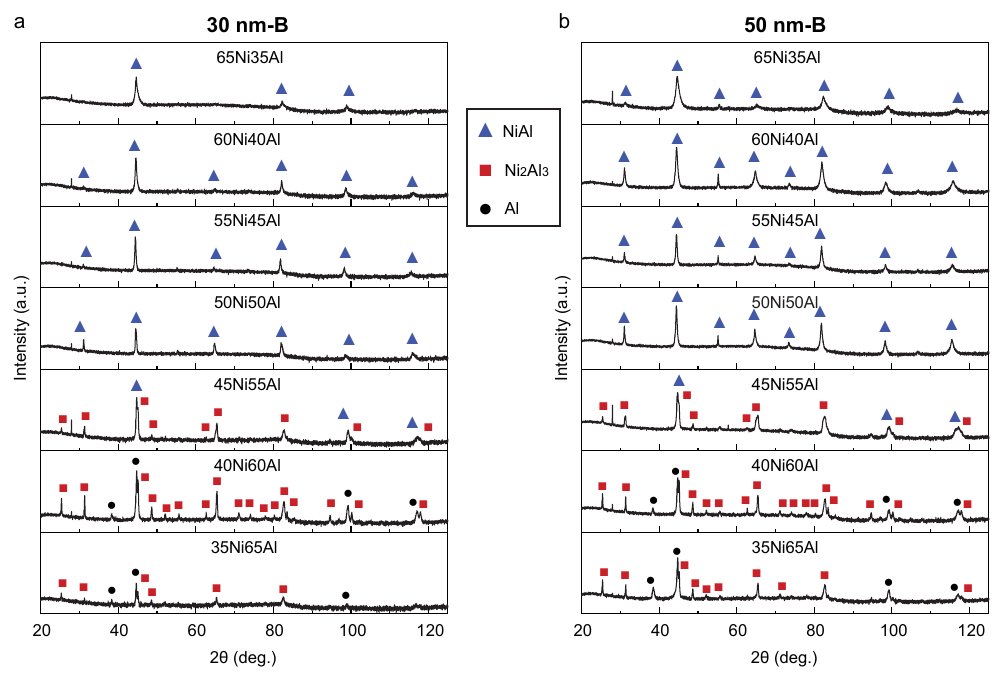}
\caption{\label{fig:XRD_ign} XRD spectra of the ignited Ni/Al RM with varying bilayer thicknesses and compositions. The triangle, square and circle shape label refers to the B2-NiAl, \ce{Ni2Al3} and Al phase respectively. (a) \qty{30}{\nano\meter}-B systems. (b) \qty{50}{\nano\meter}-B systems. The y-axis is logarithmically scaled.}
\end{figure*}

FIG.~\ref{fig:tem_ad}(a-h) shows BF-STEM and TEM microstructures of selected compositions. The Al appears bright, whereas the Ni appears dark. A wide area of the cross section of the 50Ni50Al sample with a \qty{50}{\nano\meter}-B is shown in FIG.~\ref{fig:tem_ad}(a). The layers and interfaces are uniformly flat across the entire system, with no porosity or broken interfaces. In the magnified image of the sample, the Al layers exhibit columnar grain structures with grain sizes corresponding to the layer thickness, whereas the Ni layers contain a majority of columnar grains matching the layer thickness along with some nanocrystalline grains of smaller sizes (FIG.~\ref{fig:tem_ad}(b)). The selected area electron diffraction (SAED) pattern from this region, consistent with the XRD measurements, confirms the presence of FCC nickel and aluminum phases, with no intermetallic phases detected. Typical patterns related to polycrystalline microstructures are present, where Ni(111) and Al(111) are the prominent rings, suggesting a preferred orientation. To gain more information about the interface, a BF-TEM image was acquired at a higher magnification. The intermixed layer has a thickness of about \qty{1.5}{\nano\meter}, about 40\% thinner than that reported in previous studies~\cite{gavens2000effect,fritz2013thresholds}.
The high-magnification BF-TEM imaging reveals features in the intermixed region, as indicated by the red arrow in FIG.~\ref{fig:tem_ad}(d), which may suggest the presence of a distinct phase.
However, this observation is not confirmed by TEM diffraction or XRD analysis. This apparent phase may instead result from slight tilting of the interface which results in a mixture of Al and Ni layers visible through the thickness. This effect would also lead to an underestimation of the intermixed layer thickness. Based on the available evidence, it remains unclear whether the observed features correspond to an amorphous phase or a metastable solid solution.

In FIG.~\ref{fig:tem_ad}(e-g), three compositions of the \qty{30}{\nano\meter}-B system are shown: 35Ni65Al, 50Ni50Al and 65Ni35Al, respectively. The thicknesses of the layers in the different compositions are comparable to the nominal thicknesses listed in Table \ref{tab:table1}, demonstrating extremely good control over global film compositions. As the fraction of Ni increases, the interface becomes less distinct, likely due to smaller Ni grains within the Ni layer. The SAED in FIG.~\ref{fig:tem_ad}(h) of the 65Ni35Al sample reveals no formation of intermetallic phases, even at reduced bilayer thicknesses.

FIG.~\ref{fig:indent} illustrates the experimental and MD simulation results for E and H of as-deposited Ni/Al multilayers as a function of nickel atomic percentage. Both E and H generally increase with increasing Ni content, consistent with the higher modulus and hardness of Ni compared to Al. Notably, up to \qty{50}{\atpercent} Ni, the \qty{30}{\nano\meter}-B systems exhibit lower E and H values than the \qty{50}{\nano\meter}-B systems. However, beyond \qty{50}{\atpercent} Ni, this trend reverses, with the \qty{30}{\nano\meter}-B systems demonstrating higher E and H values, indicative of a hardening effect associated with reduced bilayer thickness~\cite{schwarz2024investigating}.
For the 30Ni70Al composition, the modulus reaches \qty{102.9}{\giga\pascal} for the \qty{30}{\nano\meter}-B system and \qty{104.8}{\giga\pascal} for the \qty{50}{\nano\meter}-B system. Both systems exhibit a non-monotonic increase in modulus with Ni concentration, peaking at \qty{145.9}{\giga\pascal} for the \qty{30}{\nano\meter}-B system and \qty{140.2}{\giga\pascal} for the \qty{50}{\nano\meter}-B system in the 70Ni30Al composition. However, a notable dip is observed at the 50Ni50Al composition in the \qty{50}{\nano\meter}-B system, consistent with MD simulations.

The H values obtained from MD simulations using a perfectly columnar grain microstructure aligns well with the experimental results for Ni content equal to or greater than \qty{50}{\atpercent}. MD simulations using a single crystal Al layer yield more comparable hardness values for compositions with Ni content between 40 and \qty{60}{\atpercent}, but fail to capture the experimentally observed trend of increased hardness with increased Ni atomic percentage.

\subsection{Reaction front and phase analysis}

The experimental and MD simulation results for reaction front speeds are plotted in FIG.~\ref{fig:speed}(a) as a function of the \qty{}{\atpercent} Ni. Multilayers that experienced quenching or did not undergo ignition are marked as having a speed of zero. A frame from the quenched reaction is shown below the speed versus \qty{}{\atpercent} Ni plot, while a representative frame from a self-sustained reaction is shown at the top of the same plot. No ignition was observed for the compositions 30Ni70Al and 70Ni30Al in both systems. All the other compositions in both systems exhibited self-sustained reactions, with the exception of 35Ni65Al with 50 nm-B, which ignited but quenched immediately. The highest reaction speed occurred in the the \qty{30}{\nano\meter}-B at \qty{21.35}{\meter\per\second} in the \qty{55}{\atpercent} Ni samples while the fastest \qty{50}{\nano\meter}-B reaction speed was \qty{18.78}{\meter\per\second} in the \qty{60}{\atpercent} Ni-B. Above these Ni atomic percentages, significant reduction in the reaction speed is observed, with no ignition occurring for the 70Ni30Al composition in both bilayer thicknesses. The reaction speeds for compositions with the lower bilayer thickness are generally higher, exhibiting an average increase of 38.6\% compared to those with the higher bilayer thickness up to a composition of \qty{55}{\atpercent} Ni. Beyond this composition, the trend reverses.
The MD simulations support these experimental trends, showing a significant decline in reaction front speed for systems with less than \qty{45}{\atpercent} Ni. When the Ni content reaches 50 and \qty{60}{\atpercent}, the front velocity increases, aligning with the experimentally observed peak at similar compositions. However, as the Ni content enters the Ni-rich region (above \qty{60}{\atpercent}), both the experimental and simulation results indicate a decline in front velocity, reaching a minimum at approximately\qty{67}{\atpercent}. Beyond this point, the simulations show a gradual recovery in front velocity.

In FIG.~\ref{fig:speed}(b) the maximum temperature at the reaction front is plotted as a function of the \qty{}{\atpercent} Ni. Samples that did not ignite or were quenched are represented by a temperature of \qty{0}{\degreeCelsius}. The same trend is observed in both bilayer systems, but contrary to the reaction speed trend, \qty{50}{\nano\meter}-B display the higher temperatures. The 55Ni45Al RM produced the highest reaction temperature for both bilayer thicknesses, with values of \qty{1550}{\degreeCelsius} and \qty{1675}{\degreeCelsius} for the \qty{30}{\nano\meter}-B and \qty{50}{\nano\meter}-B system, respectively.

After the reaction, the samples were analyzed via X-ray diffraction (XRD) to identify the newly formed product phases. FIG.~\ref{fig:XRD_ign} presents the diffractograms for various compositions at both bilayer thicknesses. The experimental data indicate that the product phases for compositions with a Ni atomic fraction of 50~\% or higher align well with the Ni-Al equilibrium phase diagram~\cite{okamoto2000phase,shao2023accurate}. In contrast, for compositions with lower Ni content, the resulting phases deviate from equilibrium predictions.
For a given composition, both the \qty{30}{\nano\meter} and \qty{50}{\nano\meter}-B systems yield identical phases. Specifically, XRD analysis reveals that Al-rich multilayers predominantly form \ce{Ni2Al3}, with traces of NiAl observed in the 45Ni55Al composition. Residual aluminum is detected in compositions of 40Ni60Al and 35Ni65Al. In contrast, multilayers with Ni content of \qty{50}{\atpercent} or higher, which undergo self-sustained reactions, transform into a single-phase product identified as NiAl.
Additionally, the \qty{50}{\nano\meter} bilayer systems exhibit more intense diffraction peaks and additional phase orientations. These features are attributed to the increased total multilayer thickness and the post-ignition products.

\begin{figure*}
\includegraphics{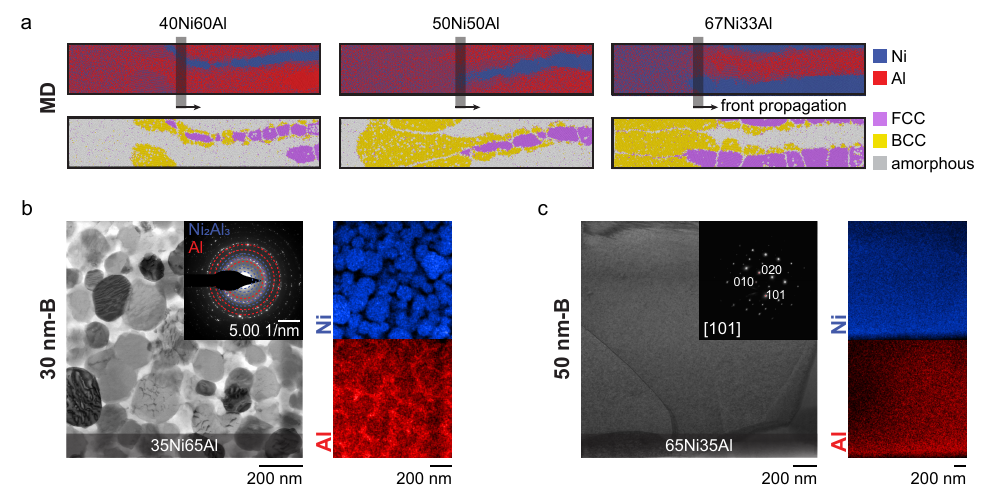}
\caption{\label{fig:post_ign_analysis} Simulated and experimental microstructures of ignited Ni/Al systems. (a) Simulated microstructures of ignited 40Ni60Al (left), 50Ni50Al (middle), and 67Ni33Al (right) systems with a bilayer thickness of \qty{15}{\nano\meter} and a \qty{2}{\nano\meter}-thick intermixed layer, observed \qty{6}{\nano\second} after ignition. The top row depicts atom types, with Ni shown in blue and Al in red, while the bottom row represents crystal structures, with FCC in purple, BCC in yellow, and amorphous regions in grey. (b) Experimental characterization of the 35Ni65Al (\qty{30}{\nano\meter}-B) system, showing a BF-STEM image, EDX composition maps, and SAED pattern with only selected dominant rings labeled, corresponding to Al and \ce{Ni2Al3}, for clarity. (c) Experimental characterization of the 65Ni35Al (\qty{50}{\nano\meter}-B) system, showing BF-STEM images, EDX composition maps, and SAED pattern confirming the presence of B2-NiAl phase.}
\end{figure*}

Figure \ref{fig:post_ign_analysis}(a) depicts MD simulation frames illustrating the diffusion process and the evolution of crystal structures for selected compositions at \qty{6}{\nano\second} post-ignition. In the 40Ni60Al composition, the excess aluminum inhibits complete crystallization into a BCC phase behind the reaction front, unlike the iso-stoichiometric system, where full crystallization is observed. Similarly, the Ni-rich 67Ni33Al system demonstrates complete intermixing, resulting in the formation of a BCC phase, likely corresponding to the B2-AlNi structure.
In contrast, the experimentally observed microstructures of two reacted samples are presented in FIG.~\ref{fig:post_ign_analysis}(b) and \ref{fig:post_ign_analysis}(c), with additional examples provided in the supplementary material (SFig.~3). The microstructures of both the 35Ni65Al sample, fabricated with \qty{30}{\nano\meter}-B (FIG.~\ref{fig:post_ign_analysis}(b)), and the 65Ni35Al sample, with \qty{50}{\nano\meter}-B (FIG.~\ref{fig:post_ign_analysis}(c)), are in agreement with MD simulation results. The 35Ni65Al sample reveals near-circular \ce{Ni2Al3} phases dispersed within an aluminum matrix, while the 65Ni35Al sample displays a microstructure dominated by large NiAl grains, with grain sizes corresponding to the total thickness of the resulting film.

\section{Discussion}
\label{sec:discussion}

\subsection{Mechanical Properties of the as-deposited samples}

The XRD analysis of the as-deposited Ni/Al multilayers reveals polycrystalline structures with distinct FCC peaks corresponding to Ni and Al, with a preferred crystallographic orientation of the Al(111) and Ni(111) planes parallel to the multilayer surface (FIG.~\ref{fig:XRDad}). A well-defined columnar morphology is observed within the Al layers, while in the Ni layers, a transition from columnar to polycrystalline structure has been observed at thicknesses exceeding \qty{20}{\nano\meter}. These structural characteristics exhibit a notable impact on the elastic and mechanical properties of the multilayers.
The experimental data indicates that the modulus generally adheres to a rule of mixtures, particularly in the \qty{30}{\nano\meter}-B system up to \qty{60}{\atpercent} Ni. However, deviations from this behavior occur at higher Ni concentrations, where the modulus no longer follows this linear trend. Similarly, hardness measurements in the \qty{30}{\nano\meter} and \qty{50}{\nano\meter}-B systems also follow the rule of mixtures, except for the Ni concentration range between 50 to \qty{60}{\atpercent}, where a drop in the values in the \qty{50}{\nano\meter}-B system are observed. We hypothesize that this behavior arises from stress relaxation when the Ni layer thickness approaches approximately \qty{20}{\nano\meter} due to a shift from constrained columnar grains to a more polycrystalline grain structure~\cite{arzt1998size}. This thickness corresponds to both the \qty{50}{\atpercent} Ni in the \qty{50}{\nano\meter}-B system and the \qty{70}{\atpercent} Ni in the \qty{30}{\nano\meter}-B system. Furthermore, a plateau in crystallite size is observed in compositions ranging from 50 to \qty{60}{\atpercent} Ni, suggesting that grain sizes depend on the thickness of the layer until \qty{20}{\nano\meter}, beyond which this dependence no longer applies (SFig.~2 in the supplementary material). The nominal thicknesses of the layers in the various multilayer systems are detailed in Table \ref{tab:table1}.
The increasing E and H values with higher Ni content can be attributed to the inherently superior mechanical properties of Ni. However, the strengthening effect correlated with reduced bilayer thickness, attributed to grain refinement mechanisms,is observed only at Ni compositions of \qty{50}{\atpercent} and higher. Below this composition, the trend reverses, which can be attributed to the presence of a thicker Ni layer where constrained Ni grains exert a more significant contribution to the overall mechanical properties.
Notably, the comparable hardness values observed in the \qty{30}{\nano\meter} and \qty{50}{\nano\meter}-B systems are unexpected, indicating that factors beyond compositional effects may influence the mechanical properties. It is hypothesized that, within this specific thickness range, the interface structure may significantly impact the mechanical response~\cite{du2023investigation}; however, further detailed investigation is required to validate this hypothesis.

While the trend between the experimental results and the MD simulation of the Young’s modulus, E, agree, an offset is observed. This offset in E is likely due to the choice of interatomic potential, which slightly underestimates the Young’s modulus of Ni~\cite{ni_interat_pot,rassoulinejad2016evaluation}.
Additionally, the presence of a Ni top layer in the experimental samples could contribute to an increased modulus. This effect may arise from the elastically deformed surface region, which likely contains a higher fraction of Ni compared to the underlying multilayer structure, thereby enhancing stiffness and increasing the measured modulus.
In some studies, it has been demonstrated that to isolate the elastic modulus of the film without significant influence from the substrate, the indentation depth must be kept below 2.5~\% of the film thickness~\cite{zak2022accurate}. However, given the limited thickness of our multilayers, adhering to this guideline is not feasible. Therefore, we have followed the conventional rule of limiting the indentation depth to less than 10\% of the total film thickness~\cite{oliver1992improved}. This approach ensures that the measured properties largely represent the film itself while minimizing substrate effects. 
The notable difference between the experimental hardness and MD simulation results can be attributed to a mismatch in microstructure, as evidenced by STEM analysis. For Ni, the experimental microstructure transitions from a columnar structure to a polycrystalline one, while the MD simulations consistently feature a polycrystalline structure with grain sizes limited to a maximum of 10 nm. For Al, MD simulations show either idealized columnar grains with grain boundaries perpendicular to the interface, leading to unrealistically high strength, or a single-crystal Al layer. The latter case yields hardness values that are more closely aligned with the experimental results. This improved agreement likely arises because the single-crystal configuration better replicates the load distribution and deformation mechanisms that occur during experimental nanoindentation. The difference between MD and experimental hardness results is unlikely to be due to strain rate effects, as yield strength has been shown to be strain-rate independent, with MD results aligning well with experimental data from the literature across a wide range of bilayer heights~\cite{PhdThesis_Fabian_Schwarz,schwarz2024investigating}. Instead, this discrepancy likely reflects significant changes in microstructure driven by variations in the Ni and Al layer thicknesses within the bilayers.
These findings highlight the key role of microstructure and layer design in the mechanical properties of multilayer systems, emphasizing the importance of integrating MD simulations with experimental data to optimize material performance.

\subsection{Reaction front and phase analysis}

Following the discussion on the as-deposited mechanical properties, we now focus on the behavior of the reaction front and phase evolution.
The observed asymmetry in reaction speed and temperature across different compositions can be attributed to the differing melting points of Ni and Al. Aluminum, with its lower melting point, melts first, facilitating favorable intermixing with Ni~\cite{rogachev2012self}. However, when there is an excess of Ni, complete intermixing does not occur, leading to inefficient energy use as the remaining Ni requires additional heat to melt. This process slows down the reaction, increases heat loss, and can eventually lead to quenching, thereby reducing the overall reaction speed and peak temperature.
The maximum front velocity was observed at a Ni content of \qty{55}{\atpercent} for the \qty{30}{\nano\meter}-B samples and \qty{60}{\atpercent} for the \qty{50}{\nano\meter}-B samples. These velocities are primarily governed by the diffusion distance and the heat of mixing. Although the heat of mixing is lower at \qty{60}{\atpercent} Ni compared to the composition at \qty{50}{\atpercent} Ni~\cite{chrifi2004enthalpies}, the diffusion distance in the \qty{50}{\nano\meter}-B system, particularly the maximum of the layer thicknesses ($h_{\mathrm{Ni}}$, $h_{\mathrm{Al}}$), is minimized. This minimization of the diffusion distance aligns with the predictions made by Hardt \textit{et al.}\cite{hardt1976effect}, who proposed that the front velocity is inversely related to the maximum diffusion distance:
\[v \sim \frac{1}{a_0 \sqrt{1 + \frac{b_0}{a_0}}}\]
where $a_{\mathrm{0}}$ and $b_{\mathrm{0}}$ correspond to the maximum and minimum of the layer thicknesses, respectively. This equation can be interpreted as:
\[v \sim \frac{1}{h_{\text{max}}}.\]
Additionally, the transition of the Ni layer to a polycrystalline structure, associated with smaller grains and higher grain boundary density, enhances diffusion and promotes faster intermixing~\cite{schwarz2021_microstructure}. This microstructural effect provides an additional explanation for the higher reaction speed observed at a composition of \qty{60}{\atpercent} Ni in the \qty{50}{\nano\meter}-B samples compared to the \qty{30}{\nano\meter}-B samples.
When analyzing the reaction rates across the two studied bilayer thicknesses, the maximum reaction speed reveals distinct mechanisms governing the reaction kinetics. In thicker bilayer systems, minimizing diffusion distance dominates over maximizing the heat of reaction, as evidenced by the shift of the maximum front velocity to \qty{60}{\atpercent} Ni. This confirms that diffusion, rather than crystallization, governs the reaction kinetics under these conditions. Conversely, in thinner bilayer systems, the influence of diffusion diminishes, with the heat of mixing becoming the primary factor driving the reaction rate. This highlights the critical interplay between diffusion and heat of mixing, which varies with bilayer thickness and composition, ultimately shaping both the reaction speed and the maximum achievable temperature.

In the MD simulations, reaction front propagation is observed even for compositions with Ni atomic fraction above 70~\%, a phenomenon primarily attributed to the absence of heat loss in these simulations. In the Ni-rich region, as the reaction front propagates, complete intermixing ceases, with Ni diffusing into the melted Al in equal quantities. From the sample 67Ni33Al in FIG.~\ref{fig:post_ign_analysis}(a-right) it can be seen that the intermixed region crystallizes into a B2-NiAl phase, while in compositions such as 75Ni25Al (SFig.~4 in the supplementary material) the remaining Ni stays in an FCC configuration. These insights provide an explanation for the increased front-propagation velocity in Ni-rich systems in MD simulations. With only a portion of the Ni layer diffusing into the liquid Al, the behavior is similar to that observed in multilayers with a 1:1 atomic ratio. However, the energy generated per unit volume decreases as a smaller fraction of the RM is transformed into the BCC NiAl phase. This results in a smaller total driving energy for the reaction front and thus decreases its velocity in films where Ni content exceeds 80~\%.
In contrast, for the Al-rich composition (Figure \ref{fig:post_ign_analysis}(a-left)), the system does not achieve complete crystallization into a BCC phase, consistent with the presence of excess aluminum atoms. Additionally, the crystallization of the intermixed layer is notably reduced compared to the iso-stoichiometric system, suggesting a weaker diffusion barrier. This reduction in the diffusion barrier properties probably contributes to the observation in MD simulations that the maximum front-propagation velocity occurs in systems with a slight excess of aluminum.
When transitioning to experimental studies, especially for compositions with low aluminum content, these results must be interpreted with caution. In cases where only a small portion of the bilayer reacts, the total energy release is significantly diminished. In MD simulations, there is no heat loss to the substrate or the surrounding environment, which allows the reaction to continue propagating. In contrast, in experimental settings, the ratio between heat loss and energy release is substantially increased, leading to quenching or conditions for no ignition, particularly in systems with high nickel content.

The experimental results from the present study exhibit strong agreement with the trends observed in MD simulations, particularly regarding phase evolution. For compositions below \qty{50}{\atpercent} Ni, the transformation predominantly results in the formation of \ce{Ni2Al3}, with compositions of \qty{40}{\atpercent} and \qty{35}{\atpercent} Ni showing an Al matrix surrounding the \ce{Ni2Al3} phase (FIG.~\ref{fig:post_ign_analysis}(a)). This correlation is especially pronounced in samples with a Ni atomic fraction of 50~\% or higher, where the formation of the B2-NiAl phase is consistently observed in both experimental and simulated results. This agreement underscores the validity of the simulations in capturing phase evolution mechanisms. However, it is crucial to consider that, unlike MD simulations, where no heat loss occurs, experimental systems experience effective cooling because of heat dissipation to the substrate and the surrounding environment. This rapid quenching, occurring on the order of milliseconds as shown in SFig.~5 in the supplementary material, significantly impacts the reaction dynamics. The increased ratio of heat loss to energy release in experiments may lead to quenching or incomplete reactions, particularly in systems with high nickel content. These cooling effects, absent in simulations, likely promote a non-adiabatic process, preventing the system from reaching equilibrium and leading to deviations from expected equilibrium phases, as observed in the Al-rich systems investigated.

\section{Conclusion}
\label{sec:conclusion}

By exploring a broader range of compositions while maintaining consistent bilayer thicknesses of \qty{30}{\nano\meter} and \qty{50}{\nano\meter}, this study expands the understanding of Ni/Al multilayer systems. Our findings demonstrate that varying the composition within this range effectively tunes both the front velocity and reaction temperature without significantly altering the mechanical properties of the material. This highlights the potential to strategically adjust composition and bilayer thickness to achieve a diverse spectrum of reaction speeds, temperatures, and mechanical characteristics.

The mechanical properties generally follow a rule of mixtures; however, deviations occur at higher nickel concentrations. Interestingly, the hardness values for both bilayer thicknesses remain unexpectedly similar, suggesting that microstructural factors or interface effects may play a more significant role than composition alone. 

Additionally, the highest front velocities were observed at different nickel contents for the two bilayer thicknesses, underscoring the interplay between the diffusion distance in thicker layers and the heat of mixing in thinner layers. This demonstrates how composition, diffusion, and reaction kinetics collectively influence reaction speed and peak temperature.

Our experimental results also reveal that composition strongly affects phase formation, with observed phases often deviating from predictions based on equilibrium phase diagrams. These deviations suggest a non-equilibrium process driven by kinetic factors such as diffusion limitations or rapid quenching. This underscores the need to consider both thermodynamic and kinetic contributions when designing and analyzing multilayer materials.

By providing a detailed analysis of how variations in composition and bilayer thickness impact reaction dynamics, microstructure, and mechanical properties, this work bridges the gap between molecular dynamics simulations and experimental results. The insights gained offer a valuable basis for the development and further optimization of advanced materials with application-specific performance requirements.

\section*{Supplementary material}
Further analysis of the as-deposited and post-ignition samples can be found in the supplementary material.

\begin{acknowledgments}
Electron microscopy analysis was performed at ScopeM, the microscopy platform of ETH Zürich. Sputter deposition of the multilayer systems was performed at FIRST, the clean room facilities of ETH Zürich. The authors are grateful to Alla Sologubenko, Henning Galinski, Pietro Tassan and Tina Curtins for their help and support.
\end{acknowledgments}

\section*{AUTHOR DECLARATIONS}
\subsection*{Conflict of Interest}
The authors have no conflicts to disclose.

\subsection*{Authors Contributions}
\textbf{N.T.}: Conceptualization (lead); Data curation (lead); Formal analysis (lead); Investigation (lead); Methodology (lead); Validation (lead); Visualization (lead); Writing – original draft (lead); Writing – review \& editing (lead). \textbf{F.S.}: Conceptualization (equal), Data curation (supporting); Formal analysis (supporting); Investigation (supporting); Methodology (equal); Visualization (supporting); Writing - Original Draft (supporting); Writing - review and editing (equal). \textbf{R.A.G.}: Data curation (supporting); Formal analysis (supporting); Investigation (supporting); Methodology (supporting); Validation (supporting). \textbf{J.G.}: Data curation (supporting); Formal analysis (supporting); Investigation (supporting); Methodology(supporting); Visualization(supporting); \textbf{R.S.}: Conceptualization (supporting);  Project administration (lead); Resources (lead); Supervision (lead); Writing–review and editing (supporting).

\section*{DATA AVAILABILITY}
The data that support the findings of this study are available from the corresponding author upon reasonable request.

%

\end{document}